\documentclass[aps,draft]{revtex4}
\usepackage[dvips]{graphicx}
\begin{document}

\title{New Kinetic Equations and Bogolyubov Energy Spectrum in a Fermi Quantum Plasma}
\author{ Nodar L.Tsintsadze and Levan N.Tsintsadze}
\affiliation{Department of Plasma Physics, E.Andronikashvili
Institute of Physics, Tbilisi, Georgia}

\date{\today}

\begin{abstract}
New type of quantum kinetic equations of the Fermi particles are derived. The Bogolyubov's type of dispersion relation, which is valid for the Bose fluid, is disclosed. Model of neutral Bose atoms in dense strongly coupled plasmas with attractive interaction is discussed. A set of fluid equations describing the quantum plasmas is obtained. Furthermore, the equation of state of a degenerate Fermi plasma is derived.
\end{abstract}


\maketitle

In recent years a huge number of works have been devoted to the investigation of
collective behavior of quantum plasmas using a set of hydrodynamic equations.
However, to the best of our knowledge, in the literature the derivation of the correct quantum fluid equations is missing.
It should be emphasized that it is not possible to derive the fluid equations with the pressure and the quantum term, which is called the Bohm's or Madelung's potential, simultaneously by the Wigner quantum kinetic equation \cite{wig}. Strictly speaking the Wigner equation is not suitable for the construction of the fluid equations.

In this Letter, we derive a new type of quantum kinetic equations of the Fermi particles
of various species. Our consideration is based on general statements of quantum mechanics, especially on the quantum statistics.
We start our investigation with a single fermi particle and for this purpose, we employ the
non-relativistic Pauli equation \cite{ber}, which reads
\begin{eqnarray}
\label{pau}
i\hbar \frac{\partial \Psi_\alpha}{\partial t}+\frac{\hbar^2}{
2m_\alpha}\Delta \Psi_\alpha -\left[ \frac{ie\hbar }{2m_\alpha c}
(\vec{A}\cdot\nabla +\nabla\vec{A})+\frac{e^2A^2}{2m_\alpha c^2}
+e_\alpha \varphi -\vec{\mu }_\alpha \cdot\vec{H}\right] \Psi _\alpha =0 \ ,
\end{eqnarray}
where $\Psi_{\alpha }=\Psi _{\alpha }(\vec{r},t,\vec{\sigma} )$ is the wave function of the single particle species $\alpha $, having the spin $\vec{s}=1/2\vec{\sigma }$ ($\sigma=\pm 1$). $\vec{A}(\vec{r},t)$
and $\varphi (\vec{r},t)$ are the vector and scalar potentials, respectively. The last
term in Eq.(\ref{pau}) is the potential energy of magnetic dipole in the external
magnetic field, the magnetic moment of which is
\begin{eqnarray}
\label{mm}
\vec{\mu}_\alpha =\frac{e\hbar }{2m_\alpha c}
\vec{\sigma }=\mu_\beta \vec{\sigma } \ ,
\end{eqnarray}
where $\mu_{\beta }$ is the Bohr magneton and $\vec{\sigma}$ is the operator of the single particle \cite{ber,lan}.

Use of the Madelung representation \cite{man} of the complex function $\Psi_\alpha$
\begin{eqnarray}
\label{mad}
\Psi_\alpha (\vec{r},t,\vec{\sigma} )=a_\alpha (\vec{r},t,\vec{\sigma} )\exp \frac{iS_\alpha (\vec{r},t,\vec{\sigma} )}{\hbar }\ ,
\end{eqnarray}
where $a_\alpha (\vec{r},t,\vec{\sigma} )$ and $S_\alpha (\vec{r},t,\vec{\sigma} )$ are real, in
the Pauli equation (\ref{pau}), yields the following two equations
\begin{eqnarray}
\label{for}
\frac{\partial a_\alpha^2(\vec{r},t,\vec{\sigma} )}{\partial t}+\nabla \cdot
\Bigl(a_\alpha^2(\vec{r},t,\vec{\sigma} )\frac{\vec{p}_\alpha }{m_\alpha }\Bigr)=0 \ ,
\end{eqnarray}
\begin{eqnarray}
\label{mom}
\frac{d\vec{p}_\alpha }{dt}=e_\alpha \Bigl(\vec{E}+\frac{\vec{v}_\alpha \times \vec{H}}{c}\Bigr)+\frac{
\hbar^2}{2m_\alpha }\nabla \frac{1}{a_\alpha }\nabla^2a_\alpha +
\mu_\beta\nabla (\vec{\sigma}\cdot\vec{H})\ .
\end{eqnarray}
Equation (\ref{for}) has an obvious physical meaning. Namely, $a_\alpha^2=|\Psi_\alpha|^2$ is the probability density of finding the single particle at some point in space with a spin $\vec{s}$. Whereas, $\vec{p}_\alpha=\nabla \vec{S}-\frac{e_\alpha}{c}\vec{A}$ is the momentum operator of the particle.
Note that if there is no spin dependence of the wave functions $\Psi_\alpha(\vec{r},t)$, then $\vec{p}_\alpha(\vec{r},t)$ becomes the ordinary momentum of the particle.
It should be emphasized that in Eq.(\ref{for})  the second, quantum Madelung, term describes the diffraction
pattern of a single electron \cite{bib}.

In the case when there are no external electric and magnetic
fields ($\vec{E}=0$, $\vec{H}=0$), after the linearization
of Eqs. (\ref{for}) and (\ref{mom}), we get the frequency of quantum oscillations of
a free electron
\begin{eqnarray}
\label{fqo}
\omega_q=\frac{\hbar k^2}{2m}\ .
\end{eqnarray}

Based on the diffraction pattern of electrons, Born has given a statistical interpretation of the wave function, which states that
in every point in space at a given time the intensity of the de Broglie waves is proportional to the probability of observing a particle at that point in space. Therefore, we shall introduce a density of probability distribution $f_S$ in phase space for the single particle
\begin{eqnarray}
\label{prob}
|\Psi |^2=\int d^3p\ f_S(\vec{r},\vec{p},t)\ .
\end{eqnarray}
This function $f_S$, obviously must satisfy the normalization condition over all phase space
\begin{eqnarray}
\label{nor}
\int d^3r\int d^3p\ f_S=\int d^3r\ |\Psi |^2=1\ .
\end{eqnarray}
Moreover, Bogolyubov has introduced a one particle distribution function \cite{bog} for the system as a whole from Liouville's theorem regarding the distribution function $f_\alpha^N(t,\tau_1,\tau_2...\tau_N),$ (where $\tau_\alpha$ is the set of coordinates and momentum components for the $\alpha$ particle), and derived the Vlasov and Boltzmann equations in the gas approximation, which means that the plasma parameter $\eta $ (representing the ratio of the average potential energy $<U>$ of particles interaction to the average kinetic energy $<\varepsilon_k>$) must be less than unity, i.e., $\eta =\frac{<U>}{<\varepsilon_k>}\ll 1$.
Note that the one particle distribution function $f(\vec{r},\vec{p},t)$ is normalized to unity, whereas the
Liouville's function $f^N$ to total number of particles, i.e., $f^N=Nf(\vec{r},\vec{p},t)$. The same relation between $f^N$ and $f(\vec{r},\vec{p},t)$ in an alternative description of kinetic theory was obtained by Klimontovitch \cite{kli,lib}. To make it more lucid, we shall give a simple explanation about a single particle and one particle distribution function \cite{lib,lan}. Namely, the probability density of the single particle is one particle per unit volume, $n^s(\vec{r},t)=|\Psi (\vec{r},t)|^2$, with dimensions 1/V. Whereas the one particle distribution function means that in spite of the large number of particles
in the unit volume all of them have only one $\vec{r}$ and $\vec{p}$. This permits us to express the number density of particles per unit volume as $n(\vec{r},t)=\int d^3p\ f(\vec{r},\vec{p},t)=N/V$. Therefore, we can write
\begin{eqnarray}
\label{ns}
n(\vec{r},t)=Nn^s(\vec{r},t)=N|\Psi (\vec{r},t)|^2=\int d^3p\ f(\vec{r},\vec{p},t)\ .
\end{eqnarray}
Thus, $n(\vec{r},t)$ is the density of quantum particles per unit volume. We
have assumed that the total number of particles of each species is conserved.

Non-equilibrium states of a Fermi quantum gas are described by the one particle
distribution function $f_\alpha (\vec{r},\vec{p},t,\vec{\sigma} ),$ which satisfies the
quantum Boltzmann equation
\begin{eqnarray}
\label{qb}
\frac{\partial f_\alpha (\vec{r},\vec{p},t,\vec{\sigma} )}{\partial t}+\left( \frac{\partial
\vec{r}}{\partial t}\cdot\nabla\right) f_\alpha (\vec{r},\vec{p},t,\vec{\sigma} )+\frac{d\vec{p}_\alpha }{dt}\frac{\partial
f_\alpha (\vec{r},\vec{p},t,\vec{\sigma} )}{\partial \vec{p}}=C(f_\alpha ) \ .
\end{eqnarray}
Equation (\ref{qb}) for quasi-particles in a Fermi liquid was written by Landau \cite{lanj,lif}.
Here $C(f_\alpha )$ is the collision integral, which describes the
variation of the distribution function due to particle collisions, the
derivative $\frac{dp_\alpha }{dt}$ is determined by the force acting on
the particle, the expression of which is given by the equation (\ref{mom}). When
the spin of particles is taken into account, the distribution function $f_\alpha $ becomes an operator with respect to the spin variables $\sigma $. In this case the total number density of particles $n_\alpha (\vec{r},t)$ equals
\begin{eqnarray}
\label{tnd}
n_\alpha (\vec{r},t)=\sum_\sigma \int \frac{d^3p}{(2\pi \hbar )^3}f_\alpha (\vec{r},\vec{p},t,\vec{\sigma})\ .
\end{eqnarray}
If there is no spin dependence of the distribution function, then $f_\alpha $ becomes the ordinary quasi-classical distribution function $f_\alpha (\vec{r},\vec{p},t)$. Note that the condition for quasi-classical motion is that the de Broglie wavelength $\hbar/p_F$ of the particle must be very small compared with the characteristic length L, over
which $f(\vec{r},\vec{p},t)$ varies considerably.

If collisions between particles were entirely negligible, each particle of
the system would constitute a closed subsystem, i.e., one could neglect the
collision integral in Eq.(\ref{qb}), and the distribution function of particles would
obey Liouville-Vlasov equation.

Hereafter, we consider the system as a spinless.
We substitute the equation of motion of single particle (\ref{mom})
(neglecting the last term) into the kinetic equation (\ref{qb}) taking into account the
definition of the density of particles (\ref{ns}) to obtain
\begin{eqnarray}
\label{tob}
\frac{\partial f_\alpha }{\partial t}+\left( \vec{v}\cdot\nabla \right) f_\alpha
+e_\alpha \Bigl(\vec{E}+\frac{\vec{v}_\alpha \times \vec{H}}{c}\Bigr)\frac{\partial f_\alpha }{
\partial \vec{p}}+\frac{\hbar^2}{2m_\alpha }\nabla \frac{1}{\sqrt{
n_\alpha }}\Delta \sqrt{n_\alpha }\ \frac{\partial f_\alpha }{\partial \vec{p}}=C(f_\alpha )\ .
\end{eqnarray}
It should be emphasized that this is a novel equation with the quantum term, which contains all the information on
the quantum effects. We specifically note also that this equation is rather simple from the mathematically point of
view.

As an example, we employ the above equation (\ref{tob}) to study the propagation of small longitudinal
perturbations ($\vec{H}=0$, $\vec{E}=-\nabla \varphi $) in an electron-ion collisionless plasmas.
For a weak field, we look for the electron and ion distribution functions in the form $f_\alpha =f_{\alpha 0}+\delta f_\alpha $, where $f_{\alpha 0}$ is the stationary isotropic homogeneous distribution function unperturbed by the field, and $\delta f_\alpha $ is the small variation in it due to field.
After linearization of Eq.(\ref{tob}) with respect to the perturbation, we assume $\delta f_\alpha $ and $\delta \varphi $ vary like $exp(\vec{k}\cdot\vec{r}-\omega t)$.

Using the Poisson's equation
\begin{eqnarray}
\label{puas}
\Delta \delta \varphi =4\pi e\left\{ 2\int \frac{d^3p}{(2\pi \hbar )^3}
\delta f_e-2\int \frac{d^3p}{(2\pi \hbar )^3}\delta f_i\right\}
\end{eqnarray}
and assuming the Fermi degeneracy temperature $T_{F}=\frac{\varepsilon_F}{K_B}$ ($K_B$ is the Boltzmann coefficient, the Fermi distribution function is the step function $f_{\alpha 0}=\Theta (\varepsilon_{F\alpha }-\varepsilon )$, where
$\varepsilon_{F\alpha }=\frac{m_\alpha v_{F\alpha }^2}{2}\ $) much more than the Fermi gas temperature,
then we obtain after some algebra the quantum dispersion equation
\begin{eqnarray}
\label{qde}
\varepsilon =1+\sum_\alpha\frac{3\omega_{p\alpha}^2}{\Gamma_\alpha k^2v_{F\alpha}^2}\left\{1-\frac{\omega }{2kv_{F\alpha}}\ln
\frac{\omega +kv_{F\alpha}}{\omega -kv_{F\alpha}}\right\}=0 \ ,
\end{eqnarray}
where
\[
\Gamma_\alpha=1+\frac{3\hbar^2k^2}{4m_\alpha v_{F\alpha}^2}\Bigl(1-\frac{\omega }{
2kv_{F\alpha}}\ln \frac{\omega +kv_{F\alpha}}{\omega -kv_{F\alpha}}\Bigr)\ ,
\]
and $\omega$ can be more or less than $kv_{Fe}$. Note that for $\omega \gg kv_{Fi},\ $ $\Gamma_i\approx 1.$

Let us first consider the electron Langmuir waves, supposing that the ion
mass $m_i\rightarrow \infty $ and $\omega \gg kv_{Fe},\ $ or the
range of fast waves, when the phase velocity exceeds the Fermi velocity of
electrons. In this case we get the dispersion relation
\begin{eqnarray}
\label{dlw}
\omega^2=\omega_{pe}^2+\frac{3k^2v_{Fe}^2}{5}+\frac{\hbar^2k^4}{4m_e^2}+...
\end{eqnarray}
which has been previously derived by Klimontovich and Silin \cite{klis}. The expression
(\ref{dlw}) exhibits that the high frequency oscillations of electrons of a degenerate
plasma remain undamped in the absence of particles collisions. Note that the
Landau damping is also absent, since according to the Fermi distribution there
are no particles with velocities greater than the Fermi velocity which could
contribute to the absorption.

As k increases, Eq.(\ref{dlw}) becomes invalid, but still $\omega > kV_{Fe}$ and the Landau damping is absent. We now introduce the Thomas-Fermi screening wave vector $k_{TF}=\frac{\sqrt{3}\omega_{pe}}{v_{Fe}}.\ $ In the limit $k^2\gg k_{FT}^2,\ $ $\omega $ tends to $kv_{Fe}\ $ (at $\ m_i\rightarrow \infty )\ $ and we obtain from (\ref{dlw})
\begin{eqnarray}
\label{klim}
\omega =kv_{Fe}\Bigl(1+2\exp \{-\frac{2(\frac{k^2}{k_{FT}^2}+
\frac{\omega_q^2}{\omega_{pe}^2})}{1+\frac{\omega_q^2}{\omega_{pe}^2}}\}\Bigr)\ .
\end{eqnarray}
If we neglect the quantum term $\omega_q$ in Eq.(\ref{klim}), then
we recover waves known as the zero sound, which are the continuation of the
electron Langmuir wave (\ref{dlw}) into the range of short wavelength. Thus the
expression (\ref{klim}) represents the quantum correction to the zero sound.

Special and very important case in a quantum plasma is an one-fluid
approximation. We further assume that the quasi neutrality
\begin{eqnarray}
\label{qneu}
n_{e}=n_{i}
\end{eqnarray}
is satisfied. This equation along with the equation of motion of ions and the
equation giving the adiabatic distribution of electrons allow us to define
the potential field. In such approximation the charge is completely
eliminated from the equations, and the Thomas-Fermi length $r_{FT}=\frac{
v_{Fe}}{\sqrt{3}\omega_{pe}}\ $ disappears with it.

In order to construct the one-fluid quantum kinetic equation, we neglect the
time derivative in the equation (\ref{tob}) of electrons, as well as the collision terms, suppose
$\vec{E}=-\nabla \varphi$ and $\vec{H}=0$,
and write the dynamic equations for the quasi-neutral plasma (\ref{qneu})
\begin{eqnarray}
\label{deq1}
(v\cdot\nabla )f_e+\nabla \Bigl(e\varphi +\frac{\hbar^2}{2m_e}\frac{1}{\sqrt{n}}
\Delta \sqrt{n}\Bigr)\frac{\partial f_e}{\partial p}=0\ ,
\end{eqnarray}
\begin{eqnarray}
\label{deq2}
\frac{\partial f_i}{\partial t}+(v\cdot\nabla )f_i-\nabla e\varphi \ \frac{
\partial f_i}{\partial p}=0\ .
\end{eqnarray}
In Eq.(\ref{deq2}) we have neglected the quantum term as a small one. Note that the Fermi distribution function of electrons
\begin{eqnarray}
\label{fde}
f_e=\frac{1}{\exp \left\{\frac{\frac{p^2}{2m_e}-U-\mu_e}{T}\right\}+1}
\end{eqnarray}
satisfies the equation (\ref{deq1}). Here $U=e\varphi +\frac{\hbar^2}{2m_e}\frac{1}{\sqrt{n}}\Delta \sqrt{n}\ ,$
and $\mu_e$ is the chemical potential.

For the strongly degenerate electrons, i.e., $T_e\rightarrow 0$ ($\mu_e=\varepsilon_F$), the Fermi distribution function
becomes the step function
\begin{eqnarray}
\label{stf}
f_e=\Theta (\varepsilon_F+U-\frac{p^2}{2m_e})\ ,
\end{eqnarray}
which allows us to define the density of electrons ($n_e=n_i=n=2\int\frac{d^3p}{(2\pi\hbar )^3}f $)
\begin{eqnarray}
\label{del}
n=\frac{p_{Fe}^3}{2\pi^2\hbar^3}\left( 1+\frac{e\varphi +\frac{
\hbar^2}{2m_e }\frac{1}{\sqrt{n }}\Delta \sqrt{n}}{\varepsilon _{Fe}}\right)^{3/2}\ .
\end{eqnarray}

We now express $e\varphi $ from the equation (\ref{del}) and substitute it into the
kinetic equation (\ref{deq2}) to obtain
\begin{eqnarray}
\label{ofk}
\frac{\partial f}{\partial t}+(v\cdot\nabla )f-\nabla \left\{ \varepsilon
_{Fe}\left(\frac{n}{n_0}\right)^{2/3}-\frac{\hbar^2}{2m_e}\frac{1}{\sqrt{n}}
\Delta \sqrt{n}\right\} \frac{\partial f}{\partial p}=0\ .
\end{eqnarray}
This is the nonlinear kinetic equation of quantum plasma in the one-fluid approximation, which incorporates the potential energy due to degeneracy of the plasma and the Madelung potential.

To consider the propagation of small perturbations $f=f_0(\vec{p})+\delta f(\vec{r},\vec{p},t)$ and $
n=n_0+\delta n(\vec{r},t),$ we shall linearize Eq.(\ref{ofk}) with respect to the perturbations, look for a
plane wave solution as $e^{i(\vec{k}\cdot\vec{r}-\omega t)},$ and derive the
dispersion equation, which resembles the Bogolyubov's dispersion
relation in the frequency range $kv_{Fe}\gg\omega\gg kv_{Fi},$
\begin{eqnarray}
\label{bog}
\omega =k\sqrt{\frac{p_F^2}{3m_e m_i}+\frac{\hbar^2k^2}{
4m_i m_e}}=k\sqrt{\frac{2\varepsilon_{Fe}}{3m_i}+\frac{\hbar^2k^2}{4m_i m_e}}\ .
\end{eqnarray}
We specifically note here that this type of spectrum (\ref{bog}) was derived by Bogolyubov for the elementary
excitations in a quantum Bose liquid, and created the microscopic theory
of the superfluidity of liquid helium \cite{lif,bogj}.

How one can explain the similar spectrum (\ref{bog}) in the Fermi gas? In the dense
and low temperature plasma the formation of bound states is possible due to the attractive character of the Coulomb force
\cite{lan,lans,pin,kre}. As is well known, when the density of particles increases and the temperature goes to zero,
the nuclear reaction leads to the capture of electrons by nuclei. In such
reaction the charge on the ions (nucleus) decreases. Because we have assumed
the quasi-neutrality, we have therefore supposed that all electrons are in the bound state with ions.
Thus one can say that in the one-fluid approximation the Fermi
plasma may become the Bose system due to the bound state or this approximation may imply the formation of the Bose atoms.

The same Bogolyubov's type of spectrum (\ref{bog}) follows from Eq.(\ref{qde}) in the range of
intermediate phase velocities $kv_{Fe}\gg\omega\gg kv_{Fi}.$ In this case, except the expression for the real part of frequency (\ref{bog}), we get the imaginary part of $\omega $ from the dispersion relation (\ref{qde})
\begin{eqnarray}
\label{im}
Im\omega =-\frac{\pi }{12}k\ \frac{p_{Fe}}{m_i} \ ,
\end{eqnarray}
which is much less than $Re\omega .$ The expression (\ref{im}) indicates that in
the absorption of oscillations the electrons play role, since their random
velocities greatly exceed the phase velocity. So that the damping rate
is determined by the electrons alone.

We shall now derive a set of fluid equations. The Boltzmann and Vlasov type of quantum kinetic equations (\ref{tob}) and (\ref{ofk}) give a microscopic description of the way in which the state of the plasma varies
with time. It is also well known how the kinetic equation can be
converted into the usual equations of fluids. Following the standard method, we can derive the equations of continuity  and motion of macroscopic quantities from Eq.(\ref{ofk})
\begin{eqnarray}
\label{con}
\frac{\partial n}{\partial t}+\nabla\cdot(n\vec{u})=0
\end{eqnarray}
\begin{eqnarray}
\label{mot}
\frac{\partial \vec{u}}{\partial t}+(\vec{u}\cdot \nabla )\vec{u}=-\frac{
K_{B}T_{Fe}}{m_i}\nabla (\frac{n}{n_0})^{2/3}+\frac{\hbar^2}{
2m_e m_i}\nabla \frac{1}{\sqrt{n}}\Delta \sqrt{n} \ ,
\end{eqnarray}
where $u(\vec{r},t)$ is the macroscopic velocity of the plasma
\begin{eqnarray}
\label{mvel}
\vec{u}=\frac{1}{n}\int \frac{2d^3p}{(2\pi \hbar )^3}\vec{v}f(\vec{r},\vec{p},t)\ .
\end{eqnarray}
Obviously from Eqs.(\ref{con}) and (\ref{mot}) after linearization follows the
Bogolyubov's type of dispersion equation (\ref{bog}).

The more general set of fluid equations for $\alpha $
kind of particles we can obtain from the equation (\ref{tob}) taking into account that any elastic scattering
should fulfill the general conservation laws of particle number, momentum and
energy
\begin{eqnarray}
\label{gcon}
\frac{\partial n_\alpha }{\partial t}+\nabla \cdot(n_\alpha \vec{u}_\alpha )=0\ ,
\end{eqnarray}
\begin{eqnarray}
\label{gmot}
\frac{\partial <\vec{p}_\alpha >}{\partial t}+(\vec{u}_\alpha \cdot
\nabla ) <\vec{p}_\alpha >=e_\alpha \Bigl( \vec{E}+\frac{1}{c}\vec{u}_\alpha\times\vec{B}\Bigr) -
\frac{1}{n_\alpha }\nabla P_\alpha +
\frac{\hbar^2}{2m_\alpha }\nabla \frac{1}{\sqrt{n_\alpha }}\Delta
\sqrt{n_\alpha }  \nonumber \\
+\frac{1}{n_\alpha }\int \frac{2dp}{(2\pi \hbar )^3}\ p_\alpha\ C(f_\alpha)\ ,
\end{eqnarray}
where $P_\alpha=\frac{1}{3m_\alpha}\int\frac{d^3p}{(2\pi\hbar)^3}\frac{p_\alpha^2}{exp\{\frac{\varepsilon_\alpha-\mu_\alpha}{
T_\alpha}\}+1}\ .$

Moreover, the equation of state of a degenerate Fermi plasma can be derived from the
kinetic equation (\ref{tob}) by multiplying it by $\frac{p_\alpha^2}{
2m_\alpha },$ integrating over the momentum and employing the equations of
continuity (\ref{gcon}) and motion (\ref{gmot}). In the non-relativistic limit the result is
\begin{eqnarray}
\label{state}
\frac{d}{dt}\ln \frac{<\varepsilon_\alpha >}{n_\alpha^{2/3}}=\frac{1
}{n_\alpha <\varepsilon_\alpha >}\int \frac{2d^3p}{(2\pi
\hbar )^3}\ \varepsilon_\alpha \ C(f_\alpha )\ ,
\end{eqnarray}
where $\varepsilon_\alpha =\frac{(p_\alpha-<p_\alpha >)^2}{2m_\alpha }$ and $
<\varepsilon_\alpha >=\frac{1}{n_\alpha }\int \frac{2d^3p}{
(2\pi \hbar )^3}\ \varepsilon_\alpha f_\alpha $ is the internal
energy of particles, and the collision terms can be positive or negative.

To summarize, we have obtained a novel kinetic equation for the Fermi quantum plasma.
In our formulation of kinetic, as well as hydrodynamic equations the quantum term, which contains all the information on
the quantum effects, is incorporated. It should be noted that the above kinetic equation has mathematical beauty. This equation was used to study the propagation of small longitudinal perturbations in an electron-ion collisionless plasmas, deriving a quantum dispersion equation. We have discussed the quantum correction to the zero sound.
For the special interest case we have derived a quantum kinetic equation in the one-fluid approximation and have shown that, when a charge and the Thomas-Fermi screening length disappear from the equation, the solution of a linear kinetic equation leads to the
Bogolyubov's type of dispersion equation, which is valid in the Bose fluid. It is clear that in the dense strongly coupled plasmas with attractive interaction the formation of the bound state is possible, so that a neutral Bose atoms can be created. A general
set of fluids equations, describing the quantum plasma was obtained from the new quantum kinetic equation.
Finally, we have derived a equation of state of a degenerate Fermi plasma in the non-relativistic limit.
These investigations may play an essential role for the description of complex phenomena that appear
in dense astrophysical objects, as well as in the next generation intense laser-solid density plasma experiments.

\end{document}